\documentclass[aps,prl,twocolumn,showpacs]{revtex4}
\usepackage{epsfig}
\usepackage{amsmath}
\usepackage{amssymb}
\usepackage{color}

\begin{document}

\title{Mechanical Properties of Transcription}
\author{Stuart A. Sevier,$^*$ }
\affiliation{Department of Physics and Astronomy, Center for Theoretical Biological Physics, Rice University, Houston, TX 77005, U.S.A.}

\author{Herbert Levine}
\affiliation{Department of Bioengineering, Center for Theoretical Biological Physics, Rice University, Houston, TX 77005, U.S.A.}
\date{\today}

\begin{abstract}
The mechanical properties of transcription have recently been shown to play a central role in gene expression. However, a full physical characterization of this central biological process is lacking. In this letter we introduce a simple description of the basic physical elements of transcription where RNA elongation, RNA polymerase rotation and DNA supercoiling are coupled. The resulting framework describes the relative amount of RNA polymerase rotation and DNA supercoiling that occurs during RNA elongation. Asymptotic behavior is derived and can be used to experimentally extract unknown mechanical parameters of transcription. Mechanical limits to transcription are incorporated through the addition of a DNA supercoiling dependent RNA polymerase velocity. This addition can lead to transcriptional stalling and important implications for gene expression, chromatin structure and genome organization are discussed. 

\end{abstract}

\maketitle

\begin{figure}[b]
\includegraphics[width=0.9\linewidth,clip=]{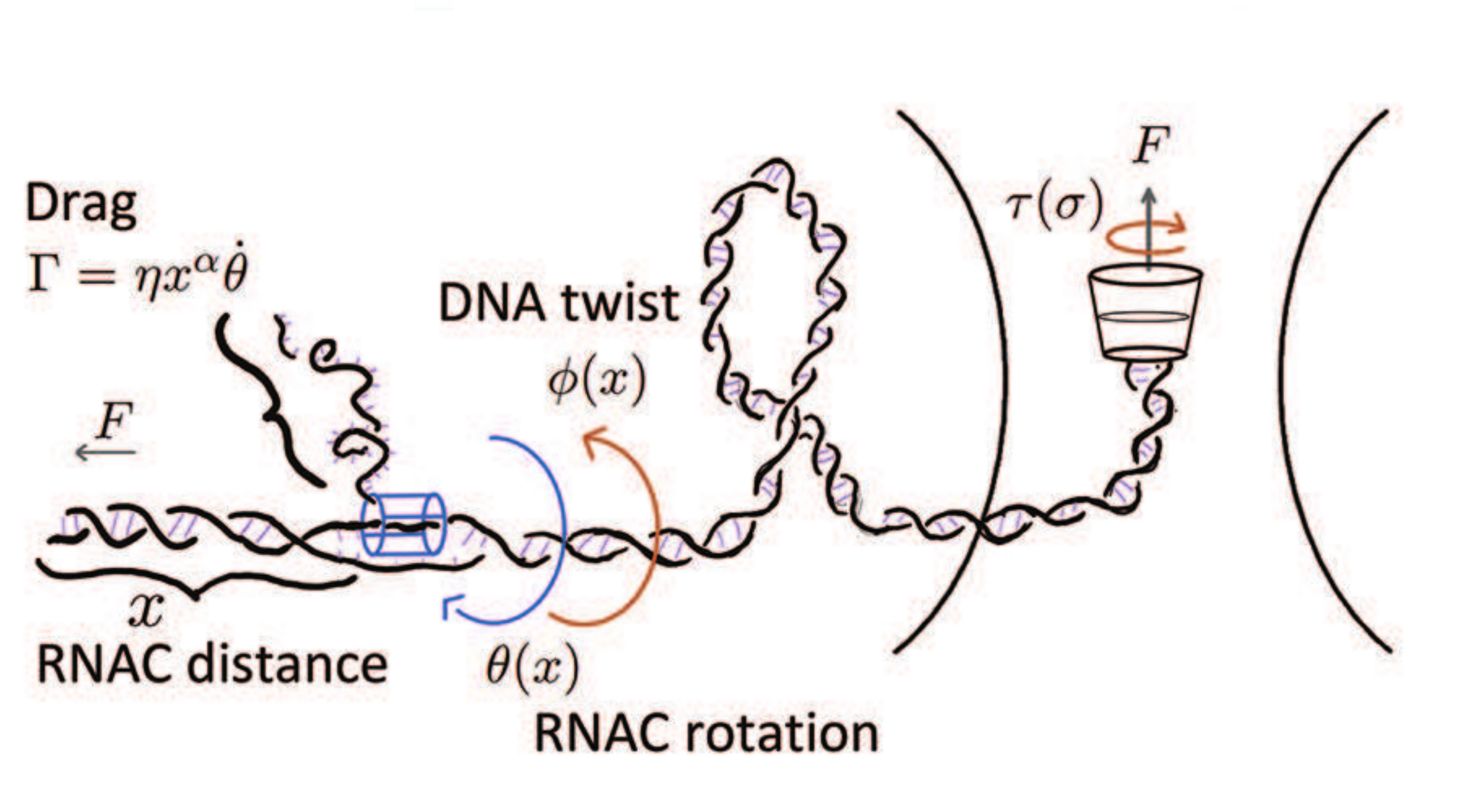}
\caption{(color online) A cartoon depicting RNA elongation $x$ through shared RNAP $\theta$ and DNA $\phi$ rotation. The DNA is attached to an optical bead for mechanical manipulation. }
\end{figure}
The helical nature of DNA introduces a physical dimension to many important biological processes. Most notably transcription, which is the first step in the conversion of genetic material into biological matter. Though the study of transcription has played a central role in modern molecular biology much of its physical foundation and behavior is just now being appreciated \cite{Lavelle2014}. Characterizing the physical aspects of transcription is essential for understanding many aspects of gene expression and may offer insights into many open problems in biology. 

The physical nature of transcription is conceptualized in the twin-domain model \cite{Liu1987} where it was first articulated that transcription and replication cause over-twisting and under-twisting of DNA. The over or under twisting of DNA is referred to as supercoiling (SC) and a number of experimental observations have revealed its central role in transcription \cite{Ma2016}. Recent results have pointed to SC and mechanical feedback as the source of transcriptional bursting \cite{Chong2014} and domain formation in bacteria \cite{Le2013}. However, a full description of this phenomena is still lacking. 

In this letter we will introduce a simplified description of the physical elements involved in transcription and construct a framework for understanding their mechanical properties. The three key elements are DNA rotation, RNA polymerase (RNAP) rotation and RNA elongation. Due to the helical nature of DNA, linear RNA elongation is coupled to rotational motion of both RNAP and connected nascent RNA. We will refer to RNAP and nascent RNA collectively as the RNA complex (RNAC). Naturally, the basic coordinates are the RNAC position along a particular gene from the transcription start site (TSS) $x$ and the relative rotation of the RNAC $\theta(x)$ and the DNA $\phi(x)$. These quantities are tied together as
\begin{equation}
\omega_{0} x=\phi(x)+\theta(x)
\end{equation}
where $\omega_{0}=1.85 nm^{-1}$ encodes the natural linking number of DNA. The relative difficulty in twisting the DNA (because of opposing torque) or difficulty rotating the RNAC (because of drag) determines the form of the functions $\phi(x),\theta(x)$. For most cases, where the DNA is not completely free to rotate, the relative amount of DNA twisting $\phi$ and RNAC rotation $\theta$ can be determined by the balance between DNA torque $\tau(\phi)$ and RNAC drag $\Gamma(x,\dot{\theta})$ as 
\begin{equation}
\tau(\phi)=\Gamma(x,\dot{\theta})
\end{equation}

While many mechanical properties of DNA are well characterized, the mechanical nature of the rotating RNAC is largely unknown. From early studies however it is clear that RNA elongation plays a key role \cite{Wu1988,Tsao1989}. Even though this is a critical factor, the coefficient and functional dependence of transcript length on the rotational drag $\Gamma$ are not known at this time. We will posit an RNAC viscous rotational drag which is linear in the rotation speed with a power-law dependence on the transcript length as $\Gamma=\eta x^{\alpha}\dot{\theta}$ where $\dot{\theta}$ is the angular speed of the RNAC and $\eta$ an unknown coefficient of friction. Dotted and primed marks denote derivatives with respect to time and space respectively. 

The mechanical properties of RNA polymerase itself are well characterized and it displays constant velocity \cite{Adelman2002} behavior over a wide range of torque (-20 to +12 pNnm) \cite{Ma2013}. We will therefore start by assuming a constant elongation rate. Later we will introduce ways for incorporating the mechanical limits of the RNA polymerase into the motion of the RNA complex.

\begin{figure}[t]
\includegraphics[width=0.9\linewidth,clip=]{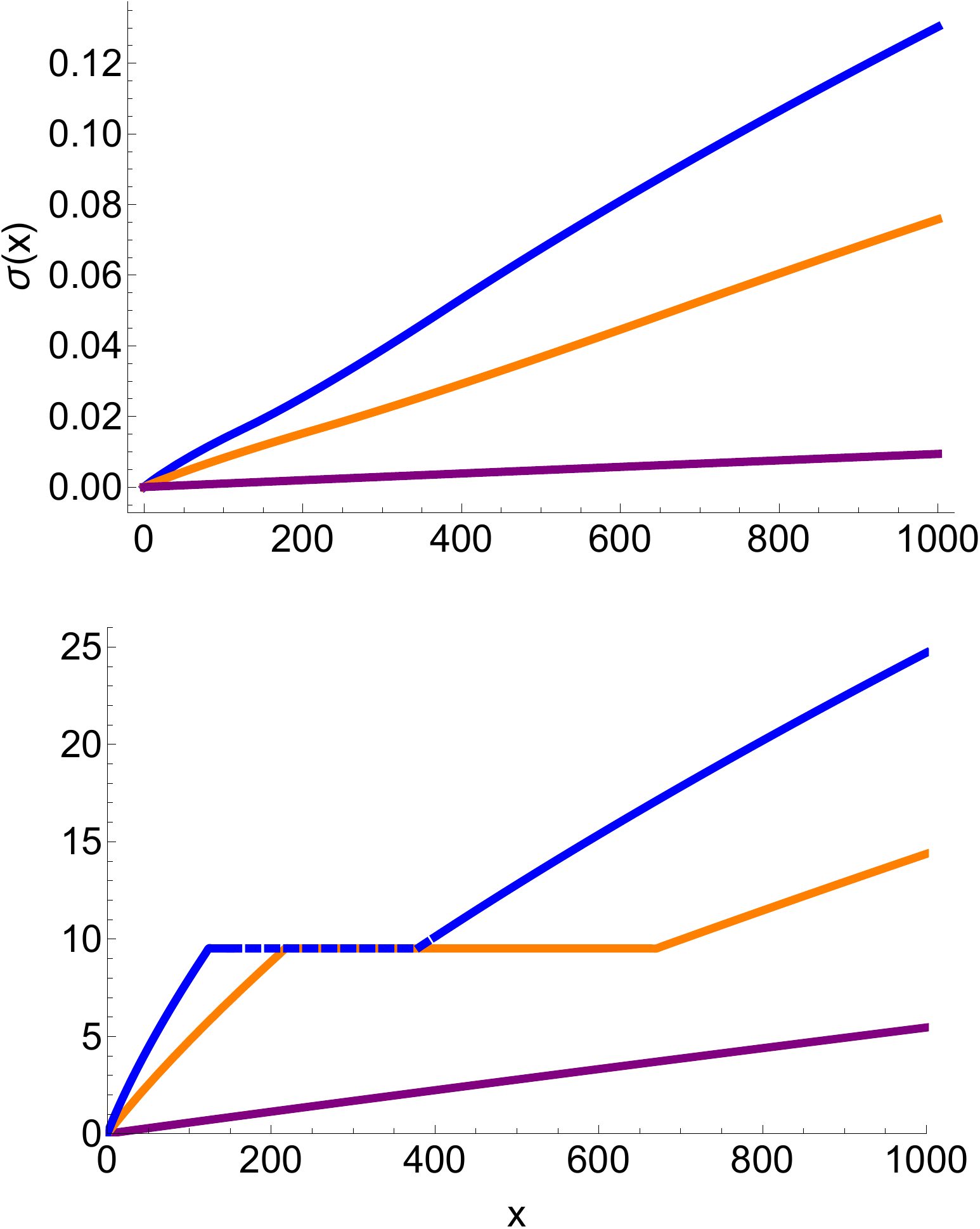}
\caption{(color online) Supercoiling $\sigma$ and torque $\tau$ as a function of RNA elongation $x$ for $L=5,10,100$ $\mu m$ (blue, orange, purple) with $\alpha=\frac{1}{2},\;\eta v=1$ at a fixed force of $F=1/2$ pNnm ($S=582.0$, $\tau_{0}=9.5$, $P=189.6$ pnNm).}\label{fig:sc_simple} 
\end{figure}

With an application of the chain rule we can turn the time derivative for RNAC rotation into a spatial derivative as $\dot{\theta}=\partial_{t}\theta(x(t))=v\theta' $ where $v=\dot{x}(t)$ is the linear velocity of the RNAC. Using this identity with equations 1 and 2 we are left with an equation of motion for DNA twist as a function of RNAP translocation
\begin{equation}
x^{\alpha} \phi'+\frac{1}{\eta v}\tau(\phi)-\omega_{0}x^{\alpha}=0
\end{equation}

We now imagine that there is an impediment to DNA twist a distance $L$ ahead of the TSS. This can be done explicitly in an in vitro experiment (shown in Fig.1) or may occur naturally for DNA in vivo with obstructions or for a closed DNA loop. Doing this turns the twist equation into an equation for supercoiling density $\sigma(x)$ (SCD) through the substitution $\sigma(x)=\frac{\phi}{\omega_{0} (L-x)}$; this expression assumes that twisting strain at the point of transcription immediately spreads throughout the specified DNA length (see below). At this time we will imagine only one barrier to DNA rotation ahead of the RNAC. These assumptions yield the equation 
\begin{equation}\label{eq:nonlsceq}
\omega_{0} x^{\alpha}(L-x) \sigma'- \omega_{0} x \sigma+\frac{1}{\eta v}\tau(\sigma)-\omega_{0}x^{\alpha}=0
\end{equation}

Under the additional assumption that the length of the gene is much smaller than the distance to the obstruction we can drop the $L^{-2}$ terms to find the SC equation 
\begin{equation}\label{eq:sceq}
x^{\alpha} \sigma'+\frac{1}{\eta v \omega_{0} L}\tau(\sigma)-\frac{x^{\alpha}}{L}=0
\end{equation}

To calculate the SCD $\sigma(x)$ as a function of translocation $x$ using eq. \ref{eq:sceq} we must specify the torque response of DNA as a function of SC $\tau(\sigma)$. Supercoiling and DNA mechanical dynamics occur on a sub-second time-scale \cite{Crut2007, VanLoenhout2012} whereas typical speeds for transcription are $10-50 \;\frac{bp}{s}$ \cite{Adelman2002}. This means that for genes on the order of $1\;kbp$ transcriptional dynamics happen on the second and minute time-scales. Additionally, RNAP operation is robust against sub-second torque fluctuations \cite{Ma2013}. Subsequently, as stated above, we expect the locally produced supercoiling at the boundary to spread throughout the allowed DNA segment on a time-scale faster than transcription occurs. More generally, we might expect to solve a supercoiling transport equation of the form
\begin{equation} \frac{\partial \Phi (\tilde{x})}{\partial t} \ = \ D \frac{\partial ^2 \Phi (\tilde{x})}{\partial {\tilde{x}}^2}
\end{equation}
with the boundary conditions for the local twist angle $\Phi = 0$ at $\tilde{x} =L$ and $\Phi = \phi (x)$ at $\tilde{x} = x$. The aforementioned limit appears if the effective rate of strain relaxation $D$ is sufficiently fast, and this will be assumed in what follows. Finally, additional sources of SC relaxation (such as topoisomerase action) will be ignored. 

Thus the torque $\tau(\sigma)$ response will be that of steady-state supercoiled DNA over a length $L$. In this framework super-coiled DNA can exist in a purely twisted, purely plectonemic or a mixed state. Following the phenomenological approach given by Marko \cite{Marko2007} the torque in a given piece of DNA held at a constant force is specified by the SCD as
\begin{equation}
\tau(\sigma)=\left\{
\begin{array}{ll}
S\sigma,\;&\sigma<\sigma^{*}_{s}\\
\tau_{0},\;&\sigma^{*}_{s}<\sigma<\sigma^{*}_{p}\\
P\sigma,\;&\sigma^{*}_{p}<\sigma\\
\end{array}
\right.
\end{equation}
where the coefficients $S,\tau_{o},P$ and SC transition values $\sigma^{*}_{s},\sigma^{*}_{p}$ are given by DNA mechanical constants and are a function of applied force. It is worth noting that the introduction of a well-defined applied force is at this time cloudy from an in vivo perspective, its experimental implementation is straight-forward. 

This formulation of $\tau(\sigma)$ yields two types of equations for supercoiling during transcription. The first is for the constant torque response in the region $\sigma^{*}_{s}<\sigma_{c}<\sigma^{*}_{p}$
\begin{equation}\label{eq:cnstsig}
x^{\alpha} \sigma_{c}'+\frac{\tilde{\tau}_{0}}{L}-\frac{x^{\alpha}}{L}=0
\end{equation}
and the second for linear torque response in the two regimes $\sigma_{l}<\sigma^{*}_{s},\;\sigma^{*}_{p}<\sigma_{l}$
\begin{equation}\label{eq:linsig}
\ x^{\alpha} \sigma_{l}'+\frac{\tilde{w}}{L}\sigma_{l}-\frac{x^{\alpha}}{L}=0
\end{equation}
where in both equations we have consolidated the constants as $\tilde{\tau}_{0}=\frac{\tau_{0}}{\omega_{0}\eta v}$ and $\tilde{w}=\frac{S}{\omega_{0}\eta v},\;\tilde{p}=\frac{P}{\omega_{0}\eta v}$ respectively. 

Analytical solutions are obtainable for both the constant and linear SC equations (eq.\ref{eq:cnstsig} and eq.\ref{eq:linsig} respectively). Numerical integration of the full non-linear equation \ref{eq:nonlsceq} (including $L^{-2}$ terms) is possible but not considered here. For the constant torque equation (eq. \ref{eq:cnstsig}) direct integration yields
\begin{equation}\label{eq:cnstsolution}
\sigma_{c}(x)=-\frac{x}{L}+\frac{\tilde{\tau}_{o}}{L}\frac{x^{1-\alpha}}{1-\alpha}+K_{1}
\end{equation}
which describes the SC as a function of RNAC elongation in the hybrid DNA response region. The constant $K_{1}$ is used to match SCD levels at the boundary.

\begin{figure}[t]
\includegraphics[width=0.95\linewidth,clip=]{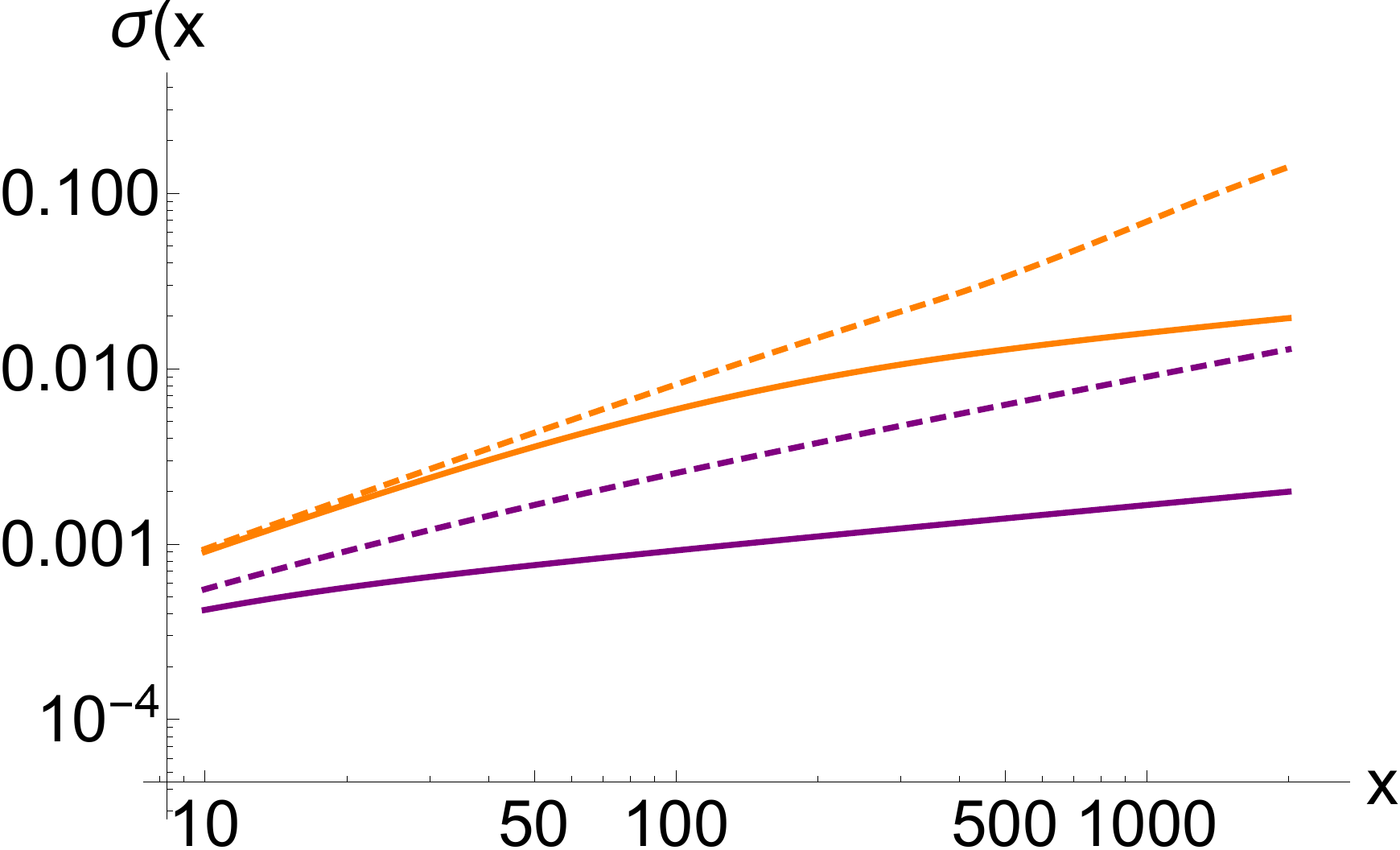}
\caption{(color online) Supercoiling $\sigma$ as a function of RNA elongation $x$ for $L=10$ $\mu m$ at a fixed force of $F=1$ pNnm ($S=622.6$, $\tau_{0}=14.5$, $P=189.6$ pnNm). The solid and dashed curves are for $\alpha=\frac{1}{2}$ and $\alpha=\frac{1}{4}$ respectively and the colors are such that $\frac{\eta_{orange}}{\eta_{purple}}=10$. The $y$ intercepts are determined by the drag coefficient $\eta$ and the slopes by the functional exponent $\alpha$. The plot illustrates how the asymptotic behavior of SC and torque would allow for direct measurements of the unknown mechanical parameters of the RNAC. }\label{fig:asympt} 
\end{figure}

The linear torque equation (eq. \ref{eq:linsig}) can be solved through the decomposition $\sigma(x)=f(x)g(x)$ with $g(x)=K_{2}e^{-\frac{\tilde{w}}{L(1-\alpha)}x^{1-\alpha}}$. This leads to the solvable equation $f'=\frac{1}{g}$ which can be used to obtain the full solution
\begin{eqnarray*}\label{eq:linearsolution}
\sigma_{l}(x)&=&\frac{K_{2}}{L}e^{-\frac{\tilde{w}}{L(1-\alpha)}x^{1-\alpha}}\\
&+&\frac{1}{L}e^{-\frac{\tilde{w}}{L(1-\alpha)}x^{1-\alpha}}\int^{x}dy \ e^{+\frac{\tilde{w}}{L(1-\alpha)}y^{1-\alpha}}
\end{eqnarray*}
for the SCD as a function of RNAC elongation in the both the pure twist and pure plectonemic regions of DNA response. Again, $K_{2}$ is a constant used to match boundary conditions. Though the solution is in the form of an integral for arbitrary $\alpha$, simple solutions exist for many rational values of $\alpha$. Typical SCD results $\sigma(x)$ and corresponding torques $\tau(\sigma)$ are shown in Fig. \ref{fig:sc_simple} for several barrier distances. 
For both the pure twisting or pure plectonemic regime, where the SC is governed by eq.\ref{eq:linsig} there is an important asymptotic response for the SCD solution $\sigma_{l}$ in the limit $x^{\alpha-1}>>\frac{\tilde{p}}{L(1-\alpha)}$ 
\begin{equation}\label{eq:asympt}
\sigma_{l} \sim \frac{1}{\tilde{w}}x^{\alpha}
\end{equation}
so that in the asymptotic limit the SC and torque directly mirror the drag associated with RNAC rotation. This is an important result as it provides an easy method for measuring the unknown RNAC drag. 
Reaching the mixed $\sigma>\sigma^{*}_{s}$ and full plectonemic $\sigma>\sigma^{*}_{p}$ regimes occur when transcription lengths exceed the values of $x^{*}_{s},\;\sigma(x^{*}_{s})=\sigma^{*}_{s}$ and $x^{*}_{p},\;\sigma(x^{*}_{p})=\sigma^{*}_{p}$ respectively. Though a closed analytical expression for $x^{*}_{s}, x^{*}_{p}$ is difficult, they can be numerically evaluated and for the case of $x^{*}_{s}$ estimated through the asymptotic expression as $x^{*}_{s}\sim(\tilde{w}\sigma^{*}_{s})^{1/\alpha}$.

Observations of the most basic elements presented in this letter can be obtained through laser trap experiments (shown schematically in Fig. 1). By observing the mechanical response of an optical bead connected to a freely rotating RNAC by a piece of DNA of length $L$, independent determination of the drag coefficient $\eta$ and the functional exponent $\alpha$ is in principle possible. This is shown in Fig. \ref{fig:asympt} where the slopes of $\sigma(x)$ on a log-log plot are determined by the value of the exponent $\alpha$ and the $y$ intercepts through the value of the drag $\eta$.

We expect the functional dependence of the drag on RNA length to reflect the effective cross-sectional area of the nascent RNA. For a random polymer the effective cross-sectional area scales as the square root of the polymer length ($\alpha=\frac{1}{2}$), however higher order RNA structure and electrostatic interactions may drive the drag away from this behavior. Additionally, it is possible that the cross-sectional area depends on angular velocity which would generate higher level terms in eq. \ref{eq:sceq}.  While neglected here such effects could be incorporated into future models. 

\begin{figure}[t]
\includegraphics[width=0.9\linewidth,clip=]{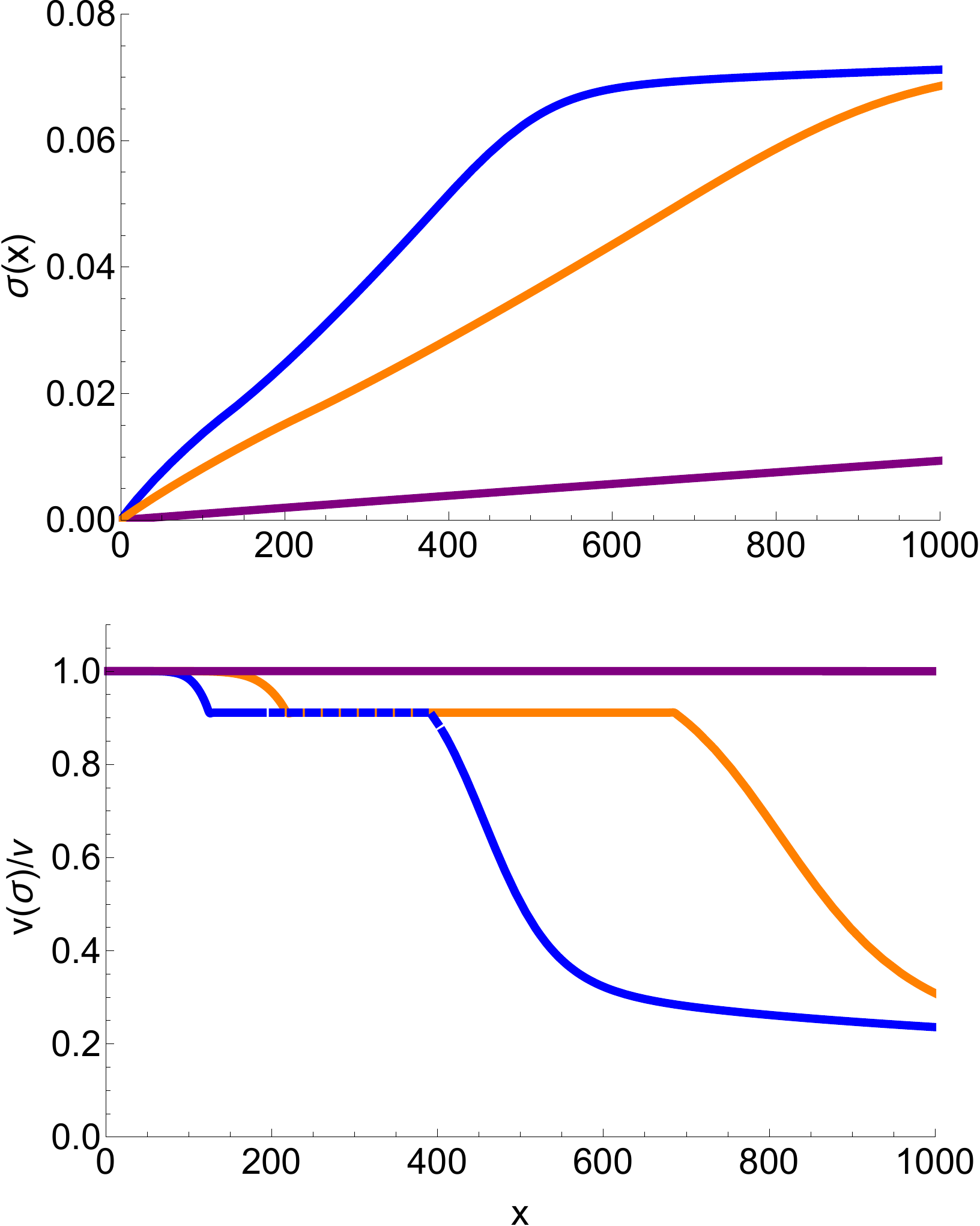}
\caption{(color online) Supercoiling $\sigma$ and velocity $v(\sigma)$ as a function of RNA elongation $x$ for $L=5,10,100$ $\mu m$ (blue, orange, purple) with $\alpha=\frac{1}{2},\;\eta v=$ at a fixed force of $F=1/2$ pNnm ($S=582.0$, $\tau_{0}=9.5$, $P=189.6$ pnNm) for a system with a torque dependent velocity. For each length the torque cutoff was set at $\tau_{c}=12 pNnm$.}\label{fig:stalled} 
\end{figure}

So far, we have assumed that RNAP moves at a constant velocity for all SC levels. However recent connections between the mechanical state of DNA, RNAC progress and transcriptional bursting have been brought to light both theoretically and experimentally \cite{Darzacq2007,Chong2014,Sevier2016}. These effects are rooted in the stalling torque of RNAP which we will call $\tau_{c}$ \cite{Ma2013}. A natural way to implement this mechanical limit is through the incorporation of a torque dependent velocity 
\begin{equation}
v(\sigma)=\frac{v_0}{1+\left(\frac{\tau(\sigma)}{\tau_c}\right)^{n}}
\end{equation}
yielding a modified equation eq.\ref{eq:sceq} which now includes a stalling component
\begin{equation}\label{eq:sceqstall}
x^{\alpha} \sigma'+\frac{1}{\eta v_{0} \omega_{0} L}\left(\tau(\sigma)+\tau_{c}\left(\frac{\tau(\sigma)}{\tau_{c}}\right)^{n+1} \right)-\frac{x^{\alpha}}{L}=0
\end{equation}
this new modified SCD equation can be numerically integrated to yield SC, torque and velocity curves as a function of elongation as shown in Fig.\ref{fig:stalled}. 

\begin{figure}[b]
\includegraphics[width=0.9\linewidth,clip=]{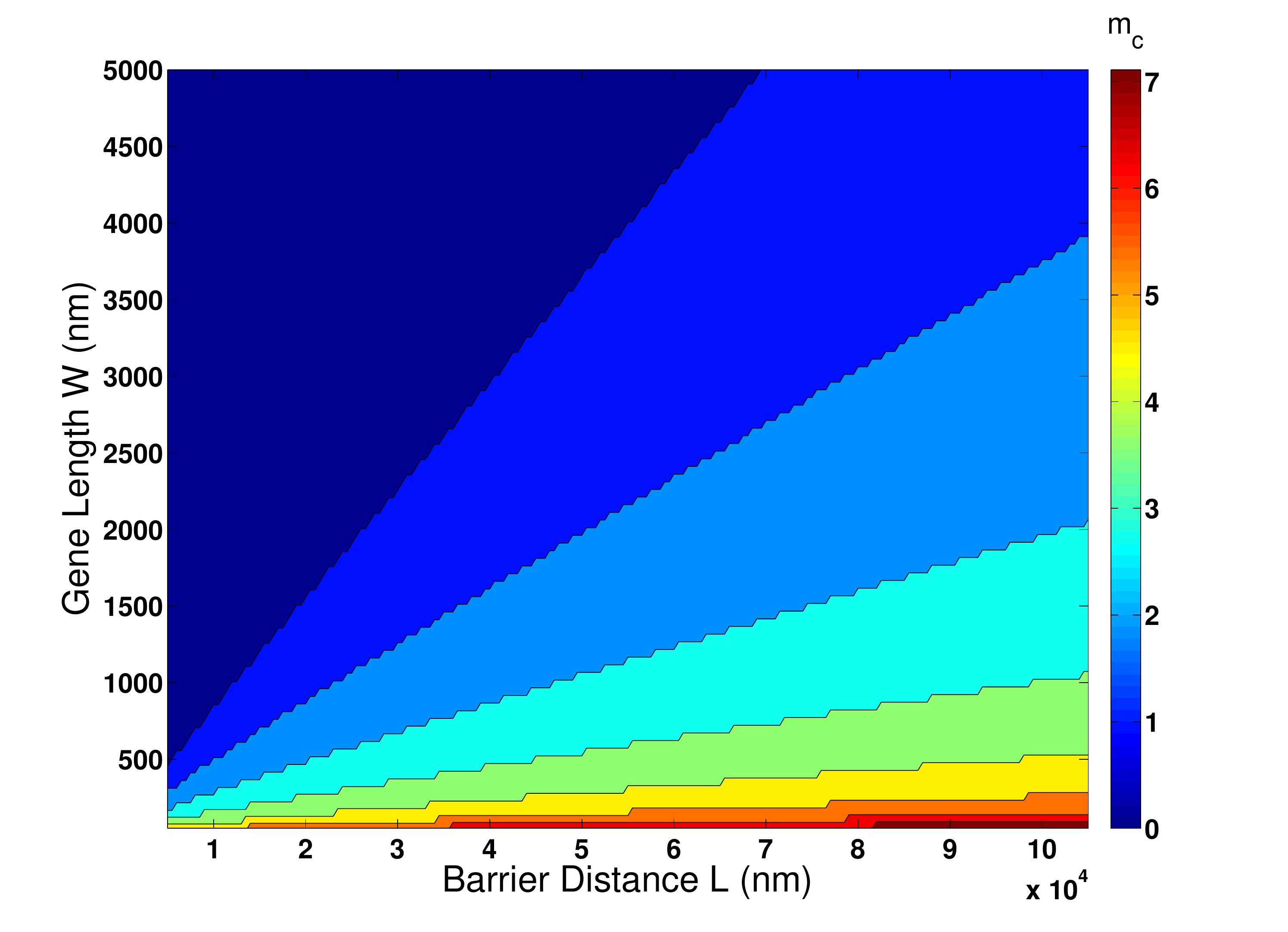}
\caption{(color online) Figure shows the number of transcripts $m_{c}$ that can be made for a gene of length $W$ against a barrier of length $L$ with no relaxation between transcription events.}\label{fig:mcestimate} 
\end{figure}

Additionally, the number of transcripts which can be made before $\tau(\sigma)>\tau_{c}$ for a gene of length $W$ given a barrier a distance $L$ is away is an important quantity for understanding the competition between mechanical frustration and relaxation in transcription \cite{Sevier2016}. To examine this we have integrated eq.\ref{eq:sceq} for various $W,L$ and prohibited relaxation between independent transcriptional events. The number of transcripts which can be made before $\tau(\sigma)>\tau_{c}$ is called $m_{c}$ and it plays an important role in bounding transcriptional noise due to RNAP stalling \cite{Sevier2016}. The dependence of $m_{c}$ on $W,L$ is shown in Fig. \ref{fig:mcestimate}.

It is thus clear that the physical location of a gene within the genome has a significant impact on its ability to be transcribed. Both the length of the gene and the distance to the barrier have significant effects on the stalling of a gene (Fig.\ref{fig:stalled}) as well as the number of transcripts that could be made without relaxation before the stalling torque is reached (Fig. \ref{fig:mcestimate}). The values shown are for convenient choices of the RNAC drag parameters and will not necessarily match real values. Experimental measurements of RNAC drag parameters (as outlined earlier in this letter) would allow for accurate predictions of the interplay between torque buildup and rotational drag. 

In addition to the mechanical state of linear DNA during transcription with a single barrier we can analyze the mechanical properties of transcription with both a forward and a backward barrier. This can be accomplished using the phenomenological approach of Marko but now with two terms of torque for a barrier a distance a distance $L_{+}$ in front of the TSS $T_{+}(\sigma_{+})$ and a barrier a distance $L_{-}$ behind the TSS $T_{-}(\sigma_{-})$. This accommodates both transcription occurring within a loop as well as transcription occurring on linear DNA with multiple barriers. For a loop the total size of the loop $L$ is simply $L=L_{+}+L_{-}$. Incorporating the torque associated with both positive and negatives supercoiling yields a modified version of eq.\ref{eq:sceq}. However, the torque formulation provided by \cite{Marko2007} can still be utilized with symmetric torque response for positive and negative SC at moderate and low forces and non-symmetric response at high forces. 

The structural conformations realized by a particular piece of DNA is constrained by the SCD. The existence of domains in bacterial DNA has been linked to transcription and has been posited to be formed out of plectonemic DNA \cite{Le2013}. The framework presented here can be used to quantitatively predict the existence of domains between actively transcribing genes. The 3D (as opposed to genomic) distance between the TSSs of two actively transcribing genes or between an actively transcribing gene and some sort of obstruction is captured by the DNA mechanical framework utilized in this letter. In the absence of adequate relaxation two actively transcribing adjacent genes with opposing orientation will generate increased SCD in the region of DNA between them. If in this region $|\sigma|>|\sigma^{*}_{s}|$ the TSSs, and the area between them, will have increased 3D proximity \cite{Marko2007}. For a piece of DNA in the regime $|\sigma|>|\sigma^{*}_{s}|$ there will be very little 3D distance between the two TSSs. Therefore, if the value of the constant torque regime $\tau_{o}$ is below that of the operational limit of the RNAP $\tau_{c}$ and $x_{g}>x^{*}_{p}$ the 3D distance between the TSSs of two active genes (or obstacle) can be point like. This behavior is a possible explanation for the existence of DNA domains in bacteria \cite{Le2013} and eukaryotes \cite{Rao2014}.

Thus, it is important to construct a framework which can accommodate simultaneously active RNACs for both the same gene as well as for neighboring genes. A model for a supercoiling dependent multi-gene system without mechanical arrest or explicit physical components of transcription has already been proposed \cite{Brackley2016}. Using elements constructed here it is possible to imagine a system where the relative rotation of multiple RNACs is determined by the mechanical state of shared DNA. The forward and a backward barrier framework outlined above addresses some of these issues. Each RNAC will operate using the basic coordinate system and torque balance outlined in equations 1 and 2. For a system of more than two genes dynamic simulations will likely be needed. A full description of a multi-gene system will benefit from measurements of the basic elements outlined in this letter and a detailed analysis of such a system is left for future work.


\acknowledgements {This work was supported by the National Science Foundation Center for Theoretical Biological Physics (Grant NSF PHY-1427654). }
\bibliographystyle{unsrt}
\bibliography{mechanical_properties}

\end{document}